\documentclass[aps,amsmath,nofootinbib,preprint,showpacs]{revtex4}
\usepackage{graphicx}
\usepackage{dcolumn}
\usepackage{bm}

\begin{document}

\title{Schwinger-Boson Mean Field Solution of Spin-1 (or 1/2) 2D Anisotropic
Heisenberg Antiferromagnet}

\author{Yanjiang JI}
\email[]{jyj@sas.ustb.edu.cn}
\author{Jinming Dong}
\affiliation{Department of Physics, Nanjing University, Nanjing
210093, People's Republic of China}
\date{06/08/1994}

\begin{abstract}
Using the Resonant Valence Bond (RVB) concept and the
Schwinger-boson mean field approximation, we investigate a two
dimensional anisotropic Heisenberg antiferromagnet. We find that
increasing the coupling ratio ($\alpha=J_{y}/J_{x}$), a
disordered-ordered phase transition appears beyond a critical
value ($\alpha_{c}$). When $\alpha>\alpha_{c}$ there exists a
finite $T_{N}$ beyond which an ordered-disordered phase transition
occurs, while for $\alpha<\alpha_{c}$ there always exists a gap.
Also we find that a decoupling temperature $T_{D}$ exists for 1D
and 2D isotropic case, when $T$ approaches $T_{D}$ the RVB order
parameter decreases rapidly to zero.
\end{abstract}


\maketitle

\section{INTRODUCTION}
Low dimensional Heisenberg antiferromagnet (HAF) system has always
been an interesting problem to theoretical physicists. Many new
concepts and techniques are developed in this field. In the 1950's
P.W.Anderson and Kubo developed the spin-wave theory
(SWT)~\cite{SWT}. Spin-wave theory predicted for the existence of
an ordered ground state and gaplessed spin-wave excitations above
it for higher dimensional magnets. In 1973~\cite{RVB},
P.W.Anderson proposed his well known Resonant Valence Bond (RVB)
model to describe some spin-1/2 systems. Then in
1983~\cite{Haldane}, Haldane proposed his conjecture, it
pronounced that the excitation spectrum of a linear-chain
Heisenberg antiferromagnet (LCHA) with integer spin has a finite
energy gap $E_{H}$ above its singlet ground state, while a LCHA
with half-odd-integer spin has a gapless spectrum. Later Affleck,
Kennedy, Lieb and Tasaki (AKLT)~\cite{AKLT} generalized the RVB
model to spin-1 LCHA by proposing the Valence-Bond-Solid (VBS)
concept to treat the conjecture. They said that these valence
bonds are formed by two 1/2 spins as a singlet $\uparrow
\downarrow  -  \downarrow \uparrow$ on the nearest neighbors,
while the two 1/2 spins on the same site should be symmetrized to
form a triplet state $S=1$. Since then many theoretical and
experimental works~\cite{Buyers,Palme,Ma} are made and now the
existence of the gap in integer spin systems is generally
accepted.

Most recently the two dimensional (2D) anisotropic Heisenberg
antiferromagnet problem has stired many people's interests. Many
theoretical techniques (spin-wave theory~\cite{Sakai},
Schwinger-boson mean field theory~\cite{Azzour} and nonlinear
$\sigma$ model~\cite{Sene}) are used to study this problem. It
seems that there exists complex phase transition in these
anisotropic systems due to the coupling ratio
($\alpha=J_{y}/J_{x}$) and temperatue ($\beta=k_{B}T$). In this
article, we will apply the Schwinger boson mean field theory
(SBMFT) to treat the problem. And a compariation to other methods
will be made. In Section II, we present the self consistent
equations and make a brief discussion. In Section III, we
calculate the self consistent equations at ground state
numerically and find a disordered-ordered phase transition in the
spin-anisotropy phase diagram. In Section IV, we discuss the
N{\rm{\'e}}el transition (AF ordered-disordered phase) due to the
temperature, and a compariation to experiments~\cite{Buyers,Palme}
on $CsNiCl_{3}$ will be made. Also, we find that a decoupling
temperature $T_{D}$ exists for 1D and 2D isotropic case, when
$T>T_{D}$ the valence bond will be decoupled, and the self
consistent equations have no solution in this region.

\section{DERIVATION OF SELF-CONSISTENT EQUATIONS}

\subsection{SBMFT Derivation}

The 2D square lattice Heisenberg antiferromagnet can be given by
the following Hamiltonian~\cite{Azzour},

\begin{equation}
H = J_x \sum\limits_{ < i,j > } {S_i  \cdot S_j }  + J_y
\sum\limits_{ < l,m > } {S_l  \cdot S_m },
\end{equation}

where the sums $<i,j>$ and $<l,m>$ are defined for the nearest
neighbors along the $x$ and the $y$ directions. $J_{x}>0$ is the
exchange coupling along the $x$ direction and $J_{y}>0$ along the
$y$ direction. Without losing the generality, we suppose the
coupling ratio $\alpha=J_y/J_x$ is not larger than 1. If we take
the limit $\alpha  \ll 1$, the problem degenerates into the 1D
case. When $\alpha$ increases, we can expect a 1D to 2D crossover.

We employ the Schwinger-boson mean field theory (SBMFT) developed
by Arovas and Auerbach~\cite{AA} to treat the Hamiltonian(1).
Unlike the Holstein-Primakoff transformation in the spin-wave
theory, we use the Schwinger-bosons to represent the spin
operator.

With the Schwingger-boson operators: $s_{\uparrow}=a$,
$s_{\downarrow}=b$, the spin $S$ can be written as:

\begin{subequations}
\begin{eqnarray}
 S^ +   = a^ +  b,S^ -   = b^ +  a \\
 S^z  = {\textstyle{1 \over 2}}\left( {a^ +  a - b^ +  b} \right) \\
 S = {\textstyle{1 \over 2}}\left( {a^ +  a + b^ +  b} \right).
\end{eqnarray}
\end{subequations}

It's easy to establish that these operators fulfill the
restriction: $[S^z,S^+]=S^+$,\\ $[S^z,S^-]=-S^-$.

Since the relation(2), we can rewrite the Hamiltonian(1) as:

\begin{equation}
H = J_x \sum\limits_{ < i,j > } {S_i  \cdot S_j }  + J_y
\sum\limits_{ < l,m > } {S_l  \cdot S_m } + \mu \sum\limits_q
{\left( {a_q^ +  a_q  + b_q^ +  b_q  - 2S} \right)}
\end{equation}

Where the index $q$ runs over the whole lattice, $a_q^ +  a_q$ and
$b_q^ +  b_q$ are the number operators of spin-up and spin-down
bosons. And $\mu$ is the lagrange multiplier to conform that there
are $2S$ Schwinger-bosons per site. Since the spin values are all
the same, the lagrange multipliers for different sites are equal.
The Hamiltonian(3) has the translational symmetry and we can apply
the Fourier transformation (FT) to it.

To simplify the discussion , let's study the $x$ direction only,
the $H_x$ term can be written as:

\begin{equation}
H_x  = J_x \sum\limits_{ < i,j > } {\left( {S_i  \cdot S_j  -
{\textstyle{1 \over 4}}n_i n_j } \right)} + J_x \sum\limits_{ <
i,j > } {{\textstyle{1 \over 4}}n_i n_j } + {\textstyle{\mu  \over
2}}\sum\limits_q {\left( {a_q^ +  a_q  + b_q^ +  b_q  - 2S}
\right)}
\end{equation}

Now let's introduce the RVB order parameter $\Delta$ (or the
valence bond for spin-1 case)~\cite{Yoshioka}: $\Delta _z  =
{\textstyle{1 \over 2}}\left\langle {b_i a_{i + z}  - a_i b_{i +
z} } \right\rangle $. $\Delta _z$ is the order parameter along
$z$($x$ or $y$)-direction, and $z$ is the unit vector along
$z$($x$ or $y$)-direction. Then the Hamiltonian has the following
form:

\begin{equation}
H_x  = C - 2\left( {{\textstyle{\mu  \over 2}} + J_x S} \right)NS
- J_x \sum\limits_{ < i,j > } {\left\{ {\left( {a_{i + x}^ +  b_i^
+   - b_{i + x}^ +  a_i^ +  } \right)\Delta _x  + h.c.} \right\}}
+ J_x \sum\limits_i {\left( {{\textstyle{\mu  \over 2}} + J_x S}
\right)n_i },
\end{equation}

where $C = 2J_x N\left| {\Delta _x } \right|^2  + J_x S^2 N$, $N$:
number of lattice sites. To simplify the derivation we ignore the
$C$ term in Hamiltonian(5), for it does not involve $\mu$. It only
shifts the whole system's energy with a certain value, and does
not influence the excitation's spectrum. Thus, the Hamiltonian(5)
can be rewritten in momentum space as:

\begin{equation}
H_x  = \sum\limits_k {\left\{ {\lambda _x \left( {a_k^ +  a_k  +
b_{ - k}^ +  b_{ - k} } \right) + \gamma _{kx} a_k^ +  b_{ - k}^ +
+ \gamma _{kx}^* a_k b_{ - k} } \right\}}  - 2NS\lambda _x
\end{equation}

Hence the Hamiltonian(3) can be expressed as:

\begin{equation}
H = H_x  + H_y  = \sum\limits_k {\left\{ {\lambda \left( {a_k^ +
a_k  + b_{ - k}^ +  b_{ - k} } \right) + \gamma _k a_k^ + b_{ -
k}^ +   + \gamma _k^* a_k b_{ - k} } \right\}}  - 2NS\lambda ,
\end{equation}

with: $\lambda _x  = {\textstyle{{\left( {\mu  + 2J_x S} \right)}
\over 2}},\lambda _y = {\textstyle{{\left( {\mu  + 2J_y S}
\right)} \over 2}},\lambda  = \lambda _x  + \lambda _y$; $\gamma
_{kx}  = 2iJ_x \Delta _x \sin k_x ,\gamma _{ky}  = 2iJ_y \Delta _y
\sin k_y ,\gamma _k = \gamma _{kx}  + \gamma _{ky}$.

Using the following Bogliubov transformation~\cite{Yoshioka}:

\begin{subequations}
\begin{eqnarray}
 a_k  = \left( {{\textstyle{{\lambda  + E_k } \over {2E_k }}}} \right)^{1/2} e^{i\theta _k /2} \alpha _k
 - \left( {{\textstyle{{\lambda  - E_k } \over {2E_k }}}} \right)^{1/2} e^{i\theta _k /2} \beta _k^ +   \\
 b_k  =  - \left( {{\textstyle{{\lambda  - E_k } \over {2E_k }}}} \right)^{1/2} e^{i\theta _k /2} \alpha _k^ +
 + \left( {{\textstyle{{\lambda  + E_k } \over {2E_k }}}} \right)^{1/2} e^{i\theta _k /2} \beta
 _k,
\end{eqnarray}
\end{subequations}

the Hamiltonian can be diagonalized as:

\begin{equation}
H_{MF}  = \sum\limits_k {\left\{ {E_k \left( {\alpha _k^ +  \alpha _k  + {\textstyle{1 \over 2}}} \right)
+ E_k \left( {\beta _k^ +  \beta _k  + {\textstyle{1 \over 2}}} \right) } \right\} - \left( {2S + 1} \right)N\lambda  }
\end{equation}

with the energy spectrum: $E_k  = \sqrt {\lambda ^2  - \gamma _k^* \gamma _k }$.

Using the diagonalized Hamiltonian(9), we obtain the free energy:

\begin{equation}
F =  - \beta ^{ - 1} \ln \left\{ {Tr\exp \left( { - \beta H }
\right)} \right\} = {\textstyle{2 \over \beta }}\sum\limits_k {\ln
\left\{ {2\sinh \left( {{\textstyle{{\beta E_k } \over 2}}}
\right)} \right\} - \left( {2S + 1} \right)N\lambda } .
\end{equation}

Thus we get the self consistent equations for $\Delta_x$ and
$\Delta_y$:

\begin{subequations}
\begin{eqnarray}
 \Delta _x  = \sum\limits_k {\left( {\coth {\textstyle{{\beta E_k } \over 2}}}
 \right)\left( {{\textstyle{{\sin k_x } \over {E_k N}}}} \right)
 \left( {J_x \Delta _x \sin k_x  + J_y \Delta _y \sin k_y } \right)}  \\
 \Delta _y  = \sum\limits_k {\left( {\coth {\textstyle{{\beta E_k } \over 2}}}
 \right)\left( {{\textstyle{{\sin k_y } \over {E_k N}}}} \right)\left( {J_x \Delta _x
 \sin k_x  + J_y \Delta _y \sin k_y } \right)}.
\end{eqnarray}
\end{subequations}

From the condition ${\textstyle{{\delta F} \over {\delta \mu }}} =
{\textstyle{{\delta F} \over {\delta \lambda }}} = 0$~\cite{AA} or
$\sum\limits_i {n_i } = 2SN$~\cite{Yoshioka}, we obtain the third
self-consistent equation: ${\textstyle{{2S + 1} \over \lambda }} =
\sum\limits_k {\left( {E_k N} \right)^{ - 1} \left( {\coth
{\textstyle{{\beta E_k } \over 2}}} \right)}$. And these
self-consistent equations can be written as integrations:

\begin{subequations}
\begin{eqnarray}
 \Delta _{\rm{x}}  = \int {\left( {{\textstyle{{dk} \over {2\pi }}}} \right)^2
 \left( {\coth {\textstyle{{\beta E_k } \over 2}}} \right)\left( {{\textstyle{{\sin k_x }
 \over {E_k }}}} \right)\left( {J_x \Delta _x \sin k_x  + J_y \Delta _y \sin k_y } \right)}  \\
 \Delta _{\rm{y}}  = \int {\left( {{\textstyle{{dk} \over {2\pi }}}} \right)^2
 \left( {\coth {\textstyle{{\beta E_k } \over 2}}} \right)\left( {{\textstyle{{\sin k_y }
 \over {E_k }}}} \right)\left( {J_x \Delta _x \sin k_x  + J_y \Delta _y \sin k_y } \right)}  \\
 {\textstyle{{2S + 1} \over \lambda }} = \int {\left( {{\textstyle{{dk} \over {2\pi }}}}
 \right)^2 \left( {E_k } \right)^{ - 1} \left( {\coth {\textstyle{{\beta E_k } \over 2}}}
 \right)}.
\end{eqnarray}
\end{subequations}

where the self consistent equations integrate from $-\pi$ to $\pi$,
and $E_k  = \sqrt {\lambda ^2  - 4\left( {J_x \Delta _x \sin k_x  + J_y \Delta _y \sin k_y } \right)^2 }$.

\subsection{Spin-spin correlation}

Since $S_i \cdot S_j$ can be expanded as:

\begin{equation}
S_i  \cdot S_j  =  - {\textstyle{1 \over 4}}\left( {a_j^ +  b_i^ +   - b_j^ +  a_i^ +  } \right)
\left( {b_i a_j  - a_i b_j } \right) + {\textstyle{1 \over 4}}\left( {a_i^ +  a_j  - b_i^ +  b_j }
\right)\left( {a_j^ +  a_i  - b_j^ +  b_i } \right),
\end{equation}

where $R_j=R_i+R$, $R=R_x+R_y$. Using expression(13), the spin-spin correlation can be expressed as~\cite{AA}:

\begin{equation}
\left\langle {S_i  \cdot S_j } \right\rangle  = \left\langle {S_0  \cdot S_R } \right\rangle
= \left| {f\left( R \right)} \right|^2  - \left| {g\left( R \right)} \right|^2 ,
\end{equation}

with:

\begin{subequations}
\begin{eqnarray}
 f\left( R \right) = {\textstyle{1 \over 2}}\int {\left( {{\textstyle{{dk} \over {2\pi }}}}
 \right)^2 e^{ik \cdot R} {\textstyle{\lambda  \over {E_k }}}\coth {\textstyle{{\beta E_k } \over 2}}}  \\
 g\left( R \right) = {\textstyle{1 \over 2}}\int {\left( {{\textstyle{{dk} \over {2\pi }}}}
 \right)^2 e^{ik \cdot R} {\textstyle{{\gamma _k } \over {E_k }}}\coth {\textstyle{{\beta E_k } \over
 2}}}.
\end{eqnarray}
\end{subequations}

Consider $R=0$, $\left\langle {S_i  \cdot S_j } \right\rangle  = \left\langle {S_i^2 } \right\rangle
= S\left( {S + 1} \right)$ ($\hbar=1$), we can renormalize the spin-spin correlation function as:
$\left\langle {S_0  \cdot S_R } \right\rangle  = F_R \left( {\left| {f\left( R \right)} \right|^2
- \left| {g\left( R \right)} \right|^2 } \right)$, $F_R$ is the renormalized factor.

\begin{table}
\begin{center}
\caption{renormalized factor $F_R$ for differrnt spin values: the SBMFT solution of large spin
system (such as spin-1) is more believable than the small spin system (spin-1).}
\begin{tabular}{|c|c|c|c|c|}
\hline  spin&   1/2&    1&  $S$&    $\infty$\\  \hline $F_R$& 3/4&
8/9& $4S(S+1)/(2S+1)^2$& 1\\
\hline
\end{tabular}
\end{center}
\end{table}

From equation(14) it's easy to conclude that if $(i,j)$ belongs to
the same sublattice $g(R)$ vanishes, and if $(i,j)$ belongs to the
different sublattice $f(R)$ vanishes. That means there exists
antiferromagnet (AF) order in this system~\cite{Azzour,AA}. When
we take the limit $R\to \infty$, both $f(R)$ and $g(R)$ vanish.
This indicates the AF order is short ranged and dereases
expotentialy~\cite{Azzour,AA}.

\section{GROUND STATE PROPERTIES}

\subsection{1D solution}

First, we shall study the $\alpha =0$ 1D limit and $\alpha =1$ 2D isotropic limit. In the 1D case the self
consistent eqs. can be written as:

\begin{subequations}
\begin{eqnarray}
 \Delta _x  = \int {\frac{{dk}}{{2\pi }}\frac{{\coth {\textstyle{{\beta E_k } \over 2}}
 J_x \Delta _x \sin ^2 k_x }}{{E_k }}}  \\
 \frac{{2S + 1}}{\lambda } = \int {\frac{{dk}}{{2\pi }}\frac{{\coth {\textstyle{{\beta E_k } \over 2}}}}{{E_k
 }}}.
\end{eqnarray}
\end{subequations}

with: $E_k  = \sqrt {\lambda ^2  - 4\left( {J_x \Delta _x \sin k_x } \right)^2 } $.

Consider the ground state $T \to 0$, i.e., $\coth {\textstyle{{\beta E_k } \over 2}} \to 1$.
We can calculate $\Delta _x$ and $\lambda$ numerically for different spin values.

\begin{table}
\begin{center}
\caption{1D SBMFT solution for different spin values at ground
state: Spin-0 case only calculated as a reference, and itself has
no specific physical meaning. Both spin-1/2 and spin-1 showed an
energy gap (2S factor in $E_H$ formula indicates that there are 2S
bosons excited for a physical excitations).}
\begin{tabular}{|c|c|c|c|}

\hline  spin&   $\Delta _x$&    $\lambda$&  $E_H=2S \omega _min$\\
\hline 0&  0&  $0.5J_x$&   0\\ \hline 1/2&    0.6792& $1.38J_x$&
$0.2425J_x$\\   \hline 1&  1.1816& $2.3647J_x$&    $0.1701J_x$\\
\hline $\infty$&   $\infty$&   $2\Delta _x J_x$&   0\\
\hline
\end{tabular}
\end{center}
\end{table}

Experiments showed that only integer spin case has a finite energy
gap (the Haldane gap). This implies that the SBMFT does not fit
for the 1D half-odd-integer spin problems. Theoretically this is
because the topological term in the long-wavelength effective
action. A formalism based on fermionic representation~\cite{AA}
for the spin operators is relevant for the half integer case.
Numerical result for spin-1 case is agree with other authors'
result. Experimently the gap ($0.41J_x$)~\cite{Ma} is larger than
the SBMFT result, this is because our model is too simple. The
model for a real spin-1 system contains a single-ion anisotropy
$D\sum\limits_i {\left( {S_i^z } \right)^2 }$.

\subsection{2D isotropic solution}

Similarly we can study the self-consistent eqs. of 2D isotropic
case~\cite{Yoshioka}. We assume $J_x  = J_y  = J$ and obtain the
2D isotropic self consistent eqs.:

\begin{subequations}
\begin{eqnarray}
 \Delta _x  = \Delta _y  = \Delta  = \int {\left( {\frac{{dk}}{{2\pi }}} \right)^2
 \frac{{\coth {\textstyle{{\beta E_k } \over 2}}J\Delta \sin k_x \left( {\sin k_x  + \sin k_y } \right)}}{{E_k }}}  \\
 \frac{{2S + 1}}{\lambda } = \int {\left( {\frac{{dk}}{{2\pi }}} \right)^2
 \frac{{\coth {\textstyle{{\beta E_k } \over 2}}}}{{E_k }}},
\end{eqnarray}
\end{subequations}

with: $E_k  = \sqrt {\lambda ^2  - 4J^2 \Delta ^2 \left( {\sin k_x  + \sin k_y } \right)^2 }$.
For the ground state $T \to 0$ and $\coth {\textstyle{{\beta E_k } \over 2}} \to 1$,
there exists an finite maximum value of $2S+1$ for the 2D integration(17).
So there exists an critical spin value $S_c$:

\begin{equation}
\left( {2S + 1} \right)_{\max }  = 2S_c  + 1 = \int {\left( {{\textstyle{{dk} \over {2\pi }}}}
\right)^2 \left( {{1 \mathord{\left/
 {\vphantom {1 {E_k }}} \right.
 \kern-\nulldelimiterspace} {E_k }}} \right)}
\end{equation}

If $S<S_c$, there exists a gap, and if $S>S_c$ gapless, a
Bose-Einstein condensation occurs at the zero energy
point~\cite{AA,Yoshioka}. Numerical calculation shows that
$S_c=0.1956$. This means that 2D isotropic Heisenberg
antiferromagnets on a square lattice with nearest coupling always
have ordered ground states, and gapless. D. Yoshioka has studied
the problem for a spin-1/2 isotropic square lattice using the
slave-fermion mean field theory. Likewise, we can generate to an
anisotropic case. Since the spectrum is gapless, the Bogoliubov
transformation(8) lost its meaning at certain $k_i$ in which
$E(k_i)=0$. D. Yoshioka introduces a special transformation to
diagonalize the mean field Hamiltonian(7) at $K_i$.

\begin{subequations}
\begin{eqnarray}
 a_{k_i }  = {\textstyle{1 \over {\sqrt 2 }}}\left( {\zeta _{k_i }  + \xi _{k_i } } \right) \\
 b_{k_i }  = {\textstyle{1 \over {\sqrt 2 }}}e^{i\theta _{k_i } } \left( {\zeta _{k_i }^ +
 - \xi _{k_i }^ +  } \right).
\end{eqnarray}
\end{subequations}

and ${\zeta _{k_i } }$, ${\xi _{k_i } }$ fulfill the following commutate relations:

\begin{subequations}
\begin{eqnarray}
 \left[ {\zeta _{k_i } ,\zeta _{k_i }^ +  } \right] = \left[ {\xi _{k_i } ,\xi _{k_i }^ +  } \right] = 1 \\
 \left[ {\zeta _{k_i } ,\xi _{k_i }^ +  } \right] = \left[ {\xi _{k_i } ,\zeta _{k_i }^ +  } \right] =
 0.
\end{eqnarray}
\end{subequations}

Hence we obtain:

\begin{subequations}
\begin{eqnarray}
 H = H\left( k \right)\left( {1 - \delta _{k,k_i } } \right) + H\left( {k_i } \right)\delta _{k,k_i }  \\
 H\left( {k_i } \right) = 2\lambda \left( {\zeta _{k_i }^ +  \zeta _{k_i }  - {\textstyle{1 \over 2}}} \right) \\
 \left[ {\zeta _{k_i } ,H } \right] = \left[ {\zeta _{k_i }^ +  ,H } \right] =
 0.
\end{eqnarray}
\end{subequations}

Now the self consistent eqs. should be rewritten as:

\begin{subequations}
\begin{eqnarray}
 2S + 1 = \frac{1}{N}\sum\limits_k {\left[ {\frac{{\lambda \coth {\textstyle{{\beta E_k } \over 2}}}}
 {{E_k }}\left( {1 - \delta _{k,k_i } } \right) + n_0 \delta _{k,k_i } } \right]}  \\
 \Delta _x  = \frac{1}{N}\sum\limits_k {\frac{{\sin k_x \left( {J_x \Delta _x \sin k_x
 + J_y \Delta _y \sin k_y } \right)}}{\lambda }\left[ {\frac{{\lambda \coth {\textstyle{{\beta E_k } \over 2}}}}
 {{E_k }}\left( {1 - \delta _{k,k_i } } \right) + n_0 \delta _{k,k_i } } \right]}  \\
 \Delta _y  = \frac{1}{N}\sum\limits_k {\frac{{\sin k_y \left( {J_x \Delta _x \sin k_x  +
 J_y \Delta _y \sin k_y } \right)}}{\lambda }\left[ {\frac{{\lambda \coth {\textstyle{{\beta E_k }
 \over 2}}}}{{E_k }}\left( {1 - \delta _{k,k_i } } \right) + n_0 \delta _{k,k_i } }
 \right]},
\end{eqnarray}
\end{subequations}

and the spin-spin correlation for the gaplessed spectrum:

\begin{subequations}
\begin{eqnarray}
\left\langle {S_0  \cdot S_R } \right\rangle  = F_R \left( {\left|
{f\left( R \right)} \right|^2
 - \left| {g\left( R \right)} \right|^2 } \right) \\
 f\left( R \right) = \frac{1}{{2N}}\sum\limits_k {e^{ik \cdot R} \left[ {\frac{{\lambda \coth
 {\textstyle{{\beta E_k } \over 2}}}}{{E_k }}\left( {1 - \delta _{k,k_i } } \right) +
 n_0 \delta _{k,k_i } } \right]}  \\
 g\left( R \right) = \frac{1}{{2N}}\sum\limits_k {e^{ik \cdot R} \frac{{\gamma _k }}
 {\lambda }\left[ {\frac{{\lambda \coth {\textstyle{{\beta E_k } \over 2}}}}{{E_k }}
 \left( {1 - \delta _{k,k_i } } \right) + n_0 \delta _{k,k_i } } \right]}
\end{eqnarray}
\end{subequations}

Compared eqs.(23) with eqs.(15), we can conclude that if there is
no gap then there is N{\rm{\'e}}el long range order, and if there
is a gap there only exists short range order and long range
disordered.

\subsection{2D anisotropic solution}

As we have discussed in section III.A there exist significant
difference between 1D and 2D isotropic Heisenberg antiferromagnet
systems. There should exist a critical coupling $\alpha _c$, when
$\alpha < \alpha _c$, the system behaves like a 1D chain (quasi 1D
spin system) and when $\alpha > \alpha _c$ it has a N{\rm{\'e}}el
order and gaplessed spectrum (2D spin system). This prediction has
been made by T.Sakai {\it et.al.} \cite{Sakai} using the spin wave
theory and M.Azzour \cite{Azzour} using the SBMFT. Recently
A.Parola {\it et.al.} \cite{Parola} declare that for spin-1/2
system there exists a disordered transition induced by anisotropy
at about $\alpha < 0.1$, they find that the disordered phase is
gapless and its wavelength can be interpreted in terms of a
decoupled 1D chains.

In M.Azzour {\it et.al.}'s article \cite{Azzour}, they provide a
solution of SBMFT for large spin approximation:

\begin{subequations}
\begin{eqnarray}
\alpha _c = S e^{-2\pi S}\\
{{\Delta _y} \over {\Delta _x}} = e^{-2\pi S}.
\end{eqnarray}
\end{subequations}

And D.S{\rm{\'e}}n{\rm{\'e}}chal {\it et.al.} \cite{Sene} work out
a nonlinear $\sigma$ model solution:

\begin{equation}
\alpha _c  = 4\exp \left[ {1 - {\textstyle{{8\pi ^2 } \over
{3\left( {a\iota } \right)^2 }}}} \right],
\end{equation}

where they choose ${a\iota }=2.1$.

Employing the condition $E_H = 0$ and $T \to 0$. We discuss the
ordered-disordered phase transition of ground state by calculating
the self consistent eqs.(12). Compared with other authors'
results, our numerical results are listed in table III. The
ordered-disordered phase diagram for different spin value and
anisotropy is given in Fig.I.

\begin{table}
\begin{center}
\caption{critical coupling $\alpha _c$ for different spin values
at ground state $T=0$.}
\begin{tabular}{|c|c|c|c|c|}

\hline  $\alpha _c$&   This article&    T.Sakai \cite{Sakai}
&  M.Azzour \cite{Azzour}&  D.S{\rm{\'e}}n{\rm{\'e}}chal \cite{Sene}\\
\hline Method&  SBMFT&  SWT&   SBMFT&  NL$\sigma$ model\\ \hline
spin-1/2& 0.13& 0.03367&
0.02&   0.55\\   \hline spin-1&  0.009& 0.0013&    0.0019&  0.03\\
\hline
\end{tabular}
\end{center}
\end{table}

\begin{figure*}
\includegraphics[clip,width=13.5cm]{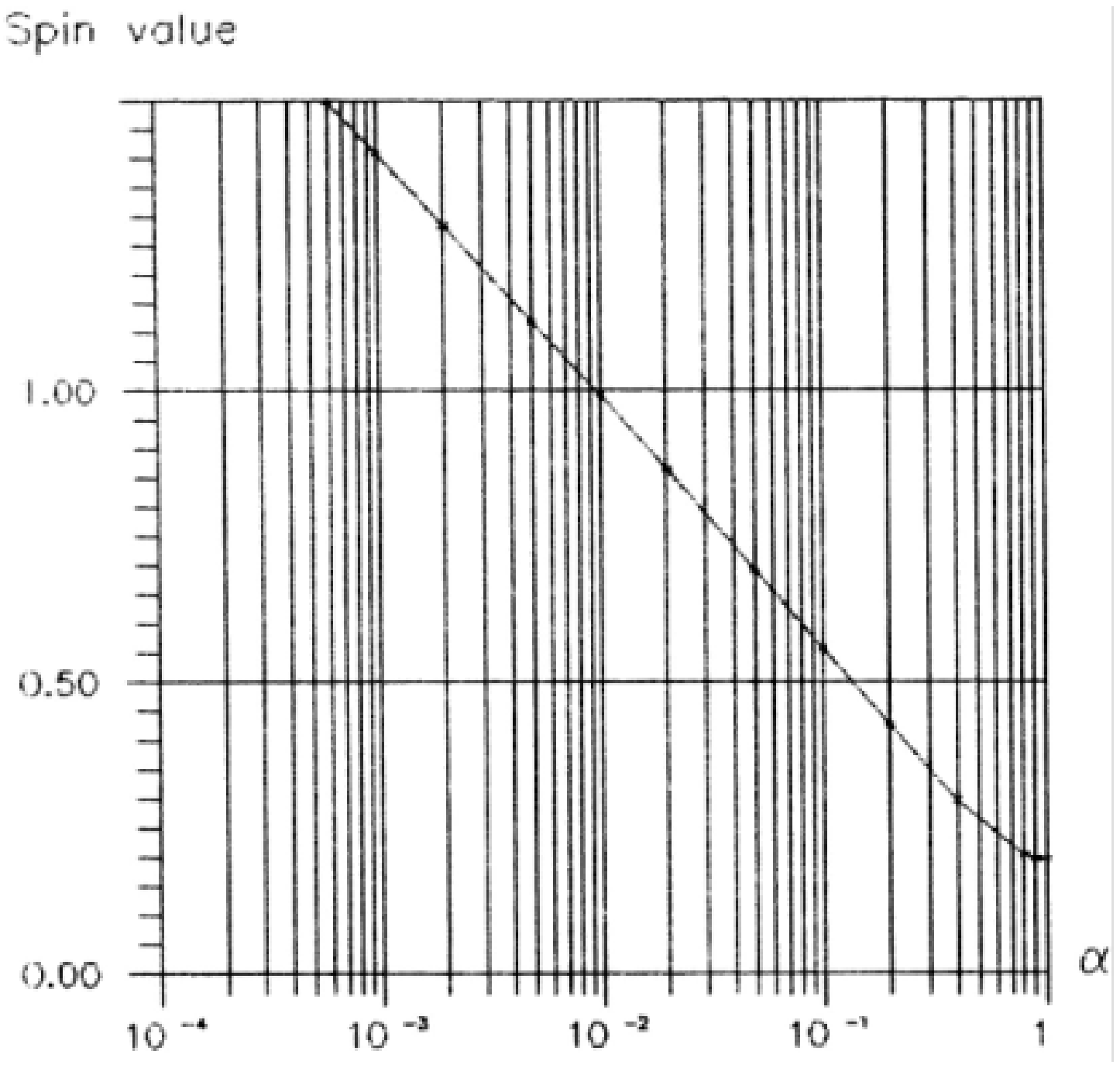}
\caption{Spin value - anisotropy $\alpha$ phase diagram at ground
state $T=0$.}
\end{figure*}

\section{EXCITED STATE PROPERTIES}

\subsection{The N{\rm{\'e}}el phase transition}

Experimentally 1D spin systems are realized in quasi-1D compounds
in which the coupling $J_x$ is much higher than the coupling
$J_y$. It is the coupling ratio $\alpha = J_y / J_x$ determines
the degree of ``quasi-one-dimensionality" of the materials. For
instance, $\alpha$ is estimated to be 0.02 in $CsNiCl_3$, and NENP
[$Ni(C_2H_8N_2)_2NO_2(ClO_4)$] 0.0006. And experiments indicate
that there exists notable difference between them. In $CsNiCl_3$
1D behavior, the existence of gap is observed above a critical
temperature (the N{\rm{\'e}}el temperature $T_N$) of about 5K
\cite{Buyers,Palme}. While in NENP 1D behavior, the existence of
gap is observed at temperature as low as can be reached. This
indicates that for spin-1 system $\alpha _c$ is not larger than
0.02. The result from nonlinear $\sigma$ model in table III seems
too large \cite{Sene}, while SBMFT solution and SWT solution are
permitted.

Calculating the self consistent eqs.(12) numerically, we can find
the $T_N$ for $CsNiCl_3$. It approximates 1K, much lower than the
experimental result. Perhaps it is due to the compound
$CsNiCl_3$'s structure is $ABX_3$ type, not cubic \cite{Palme}.
And for a real system, the Hamiltonian should include the
anisotropic term $D\left( {S_i^z } \right)^2$ \cite{Buyers,Palme}.
D is produced by the coupling of a spin to the anisotropic orbital
motion.

R.Botet {\it et.al.} \cite{Botet} found that a gap exists for $1.6
\ge {\textstyle{D \over J}} \ge  - 0.5$, and the gap decreased
rapidly for negative D. Experiments indicate that the anisotropy
constant is negative ${\textstyle{D \over J}} =  - 0.038$
\cite{Buyers}. So the calculated $T_N$ ($D=0$) should be smaller
than the $T_N$ for a real system ($D =  - 0.038J$). The
ordered-disordered phase diagram for different spin values and
temperature with $\alpha=0.02$ is given in Fig.2.

\begin{figure*}
\includegraphics[clip,width=13.5cm]{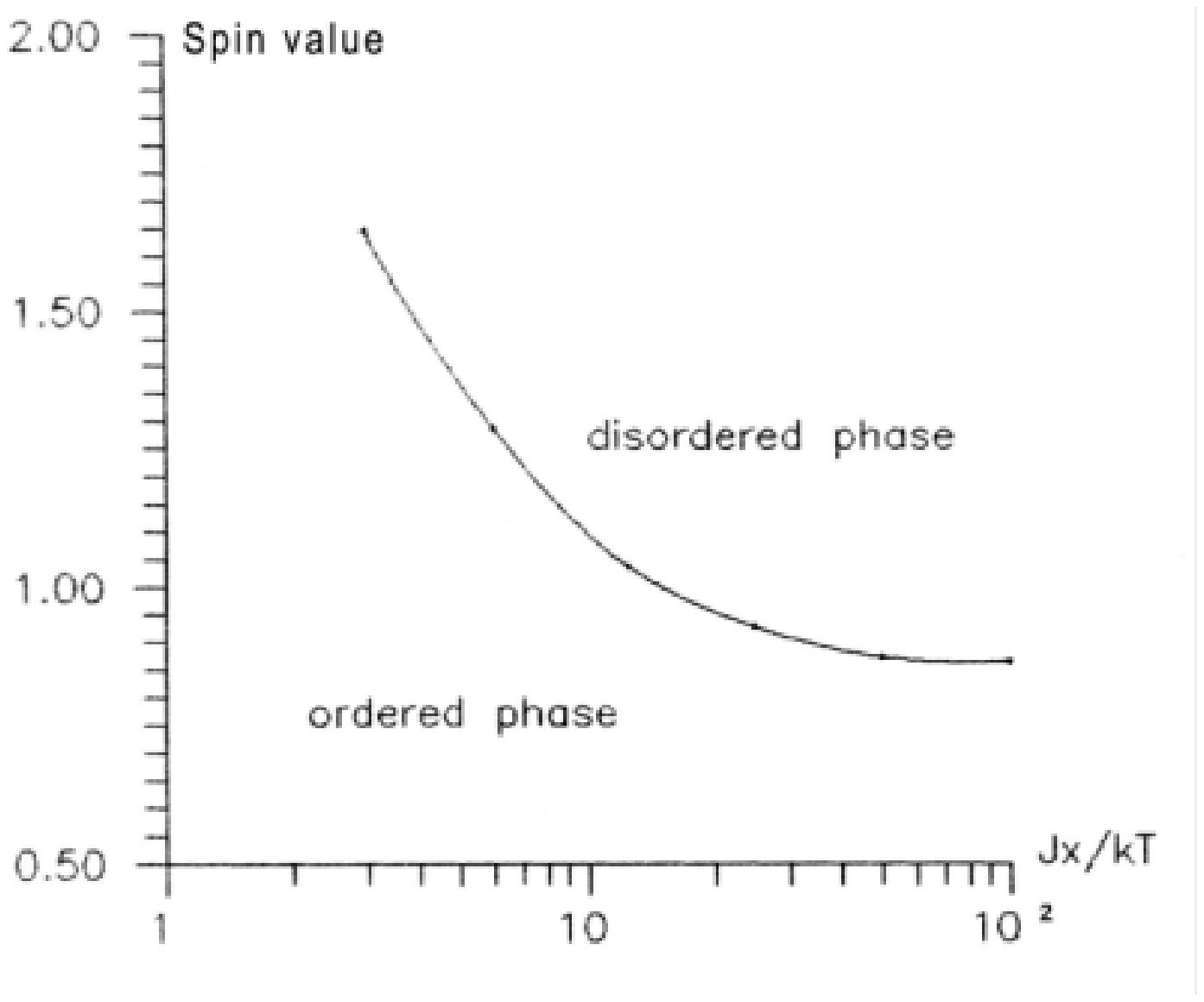}
\caption{Spin value - temperature $J_x / k_{B}T$ phase diagram
with $\alpha =0.02$.}
\end{figure*}

\subsection{The temperature-anisotropy phase diagram of spin-1 system}

Numerical solution of eqs.(12) indicate that when $\beta \to 0.46$
the RVB order parameter $\Delta$ decreases rapidly to zero for
both 1D and 2D isotropic spin-1 systems. Suppose $\Delta _x=
\Delta _y = 0$, we can calculate the critical temperature $T_D$,
denoting that when $T \ge T_D$ the boson pairs on the nearest
neighbors will be decoupled.

\begin{equation}
\coth {\textstyle{\lambda  \over {2T_D }}} = 2S + 1
\end{equation}

Solving the equation (26), we find that $T_D = 2.16$ for spin-1
case of $\alpha = 1$ and $\alpha = 0$. But, for the anisotropic
case $ 0 < \alpha  < 1 $, the self consistent eqs.(12) has no
solution under the condition $\Delta _x = \Delta _y = 0$ or
$\Delta _x \ne 0, \Delta _y = 0$. This indicates that when
increasing the temperature, the Valence-Bond-Solid concept lost
its validity before the boson pairs on the nearest sites
decoupled. The phase diagram of spin-1 Heisenberg antiferromagnet
can be divided into three regions: the ordered phase, the
disordered phase, and the decoupled phase. Valence-Bond-Solid
concept lost its validity in the decoupled phase, we must consider
other order parameters. The phase diagram for spin-1 2D
anisotropic Heisenberg antiferromagnet is given in Fig.3.

For spin-1/2 systems such as $KCuF_3$ and Cu-O layers etc., $T_D$
is calculated out to be 0.910.

\begin{figure*}
\includegraphics[clip,width=13.5cm]{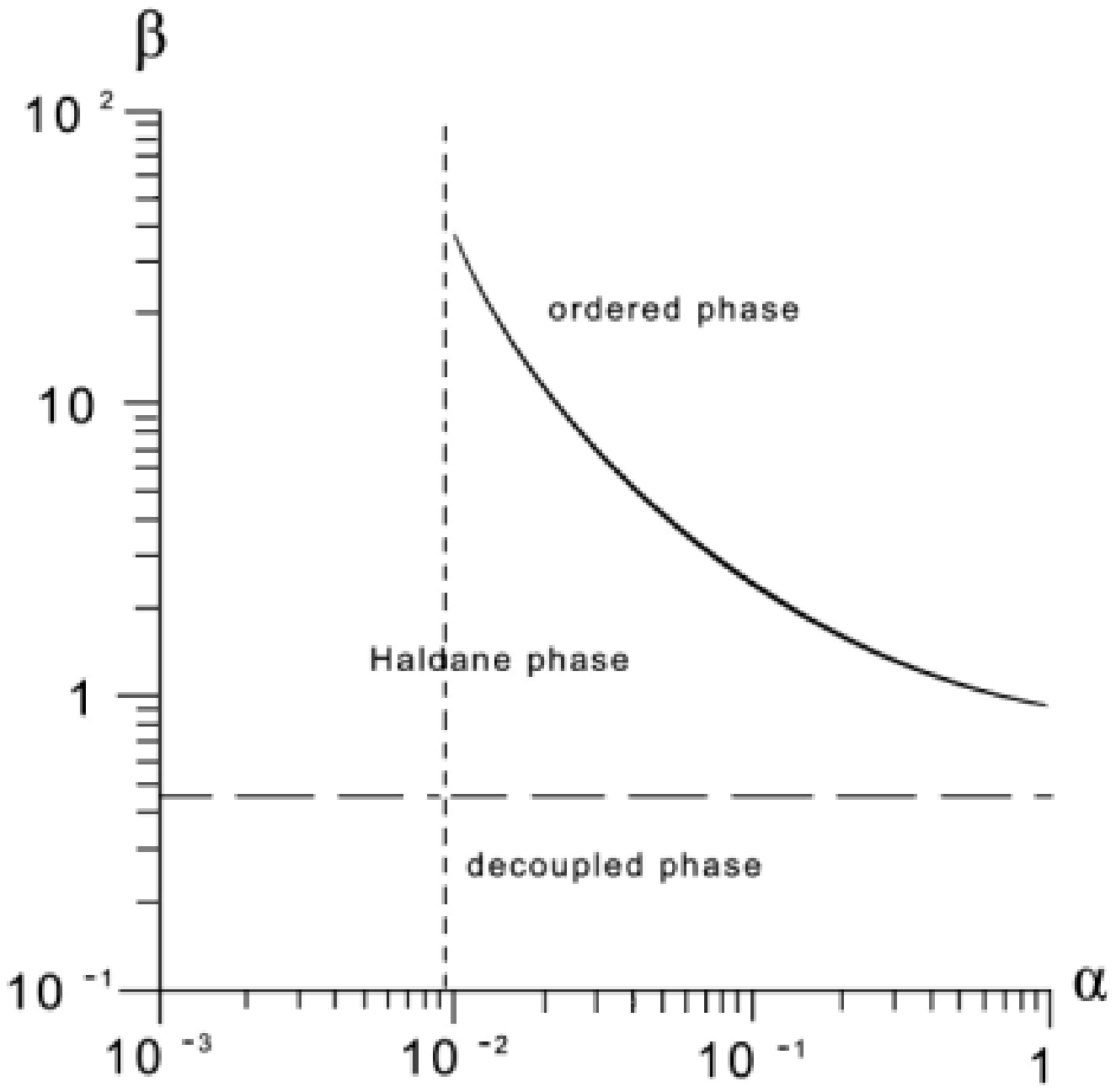}
\caption{Temperature $J_x / k_{B}T$ - anisotropy $\alpha$ phase
diagram for spin-1 2D anisotropic Heisenberg antiferromagnet.}
\end{figure*}

\section{SUMMARY}

Using the the Schwinger-boson mean field theory, we have
investigated the phase transition in anisotropic 2D Heisenberg
antiferromagnet for various spin values. We find that the
anisotropic HAF system can be characterized as two types according
to different coupling ratio $\alpha = J_y / J_x$. When $\alpha <
\alpha _c$, such as NENP for spin-1 system, it belongs to the
quasi 1D spin systems. For the quasi 1D spin systems, the
disordered phase always exists. When $\alpha > \alpha _c$, such as
$CsNiCl_3$ for spin-1 system, it belongs to the usual 2D spin
system. For the usual 2D spin system, the disordered phase only
exists when $T > T_N$. So we can use $CsNiCl_3$ and other $ABX_3$
type compounds (such as $RbNiCl_3$, $CsNiF_3$ et.al.) for testing
anisotropic 2D HAF problem and use NENP for testing Haldane
conjecture and other 1D spin-chain HAF problems. Also we find that
for the 1D and 2D isotropic HAF system there exists another
critical temperature $T_D$, when $T \ge T_D$, $\Delta _x = \Delta
_y = 0$, the boson pairs will be decoupled. But for the
anisotropic case ($0 < \alpha < 1$), there has not such a
decoupled solution for $\Delta _x = \Delta _y = 0$ or $\Delta _x
\ne 0, \Delta _y = 0$. This implies order parameters other than
RVB should be considered in the high temperature region.

\begin{acknowledgments}
Several discussions with Dr. Zheng-Hao Wang are gratefully
acknowledged.
\end{acknowledgments}

\end{document}